\documentclass[11pt]{article}
\usepackage{epsfig} 
\newcommand{\Dslash}{D \! \! \! \! /}

\newcommand{\half}{\mbox{\small{$\frac{1}{2}$}}} 
 
\newcommand{\Nf}{N_{\!f}} 
\newcommand{\MSbar}{\overline{\mbox{MS}}}

\begin{document}
\title{One loop $\MSbar$ gluon pole mass from the LCO formalism} 
\author{R.E. Browne \& J.A. Gracey, \\ Theoretical Physics Division, \\ 
Department of Mathematical Sciences, \\ University of Liverpool, \\ P.O. Box 
147, \\ Liverpool, \\ L69 3BX, \\ United Kingdom.} 
\date{} 
\maketitle 
\vspace{5cm} 
\noindent 
{\bf Abstract.} We compute the one loop corrections to the pole mass of the 
gluon in the $\MSbar$ scheme in the Landau gauge in both the Curci-Ferrari 
model and the local composite operator formalism with $\Nf$ flavours of 
massless quarks. For the latter we determine an estimate for the gluon mass
using the effective potential of a local dimension two composite operator and
find, for example, $m_{\mbox{\footnotesize{gluon}}}$ $=$ $2 \!\cdot \! 10 
\Lambda_{\mbox{\footnotesize{$\MSbar$}}}$ in Yang-Mills theory. 

\vspace{-16cm}
\hspace{13.5cm}
{\bf LTH 625}

\newpage

The issue of whether the gluon obtains a dynamically generated mass has been
a popular topic of investigation in recent years. Following the work of
\cite{1,2,3,4} who observed that the perturbative vacuum of QCD is unstable, 
one of the main activities has been on the numerical evaluation of the vacuum 
expectation value of the square of the gauge potential, $\langle \half A_\mu^2 
\rangle$. Various methods have been used to achieve this ranging from 
combinations of lattice computations with the operator product expansion and 
instanton considerations, \cite{5,6,7,8,9,10,11}, to a more theoretical 
approach of the local composite operator formalism of, for instance,
\cite{12,13,14,15}. Moreover, there is evidence from phenomenology that the 
existence of a gluon mass in the range of $500$-$800$ MeV may provide a more 
accurate explanation of various experimental data. Indeed, a valuable summary 
table of current gluon mass estimates has been given in the article by Field,
\cite{16}. Whilst the operator $\half A_\mu^2$ suffers from the immediate 
objection of not being a gauge invariant entity, it has been shown how to 
relate it to a dimension two gauge invariant physical operator, which is the 
minimization of $A_\mu^2$ over all gauge configurations \cite{12,17,18}.This 
operator, albeit non-local, reduces to a local operator in the Landau gauge and
it is solely in this gauge that, for example the lattice results of 
\cite{5,6,10,11} have been determined. Indeed the local composite operator 
(LCO) formalism of \cite{12,13} was originally developed in the Landau gauge 
but recently an estimate of $\langle \half A_\mu^2 \rangle$ has been determined
in arbitrary linear covariant gauge, \cite{17}.

Whilst there is much activity in trying to ascertain the existence of a
dynamical gluon mass, there appears to be less effort into standardizing mass 
estimates. For instance, in the quark sector of QCD the estimates of the 
various quark masses by methods such as sum rules, lattice regularization and 
the operator product expansion are all expressed as the $\MSbar$ running mass 
at the scale of $2$GeV. Although clearly measurements are not always made at 
this scale. To connect the mass estimates one requires an as accurate as 
possible evaluation in perturbation theory of the quark mass anomalous 
dimension in the $\MSbar$ scheme. This is currently available at four loops, 
\cite{19,20,21,22,23}. Moreover, the relation between the quark pole masses and
the running mass is known at three loops, \cite{24}. For the same problem for a
gluon mass the analogous quantities are not available to as high an order. For 
instance, the running of the naive gluon mass operator, 
$\half A_\mu^2$~$-$~$\alpha \bar{c} c$, in the non-linear Curci-Ferrari gauge, 
\cite{25}, is known at three loops, \cite{26}. In the Landau gauge, it 
transpires that it is not an independent renormalization being the sum of the 
gluon and ghost anomalous dimensions, which is a result that derives from a 
Slavnov-Taylor identity, \cite{27}. This has recently been exploited to obtain
the four loop running in the Landau gauge for the $SU(3)$ colour group,
\cite{28}. However, the relation between the pole mass of the gluon and the 
running gluon mass is not yet available for QCD at {\em one} loop. Therefore, 
it is the aim of this article to provide such a relation for QCD which will 
build on the Yang-Mills expression recently given in \cite{29} for the 
Curci-Ferrari gauge. Moreover, since the LCO formalism has provided estimates 
for a dynamically generated gluon mass which are comparable with other methods
we will also determine the relation for that approach as well. This will 
provide a clean estimate for a gluon mass, since in \cite{12} the effective 
potential for the operator $\half A_\mu^2$ was developed at two loops in the 
Landau gauge. However, there the estimate for a dynamical gluon mass was based 
on determining the value of an {\em effective} gluon mass which was by 
definition a {\em classical} mass. It seems to us that a more appropriate 
quantity to estimate through the effective potential approach would be a one 
loop quantity derived from the gluon two-point function such as the pole mass. 
This is the second aim of the article.  

We begin by defining our notation. We recall that the QCD Lagrangian in a 
linear covariant gauge is 
\begin{equation}
L^{\mbox{\footnotesize{QCD}}} ~=~ -~ \frac{1}{4} G_{\mu\nu}^a 
G^{a \, \mu\nu} ~-~ \frac{1}{2\alpha} (\partial^\mu A^a_\mu)^2 ~-~ 
\bar{c}^a \partial^\mu D_\mu c^a ~+~ i \bar{\psi}^{iI} \Dslash \psi^{iI} 
\label{qcdlag} 
\end{equation} 
where $\alpha$ is the gauge fixing parameter, 
$G^a_{\mu\nu}$~$=$~$\partial_\mu A^a_\nu$~$-$~$\partial_\nu A^a_\nu$~$-$
$g f^{abc} A^b_\mu A^c_\nu$ and $\psi^{iI}$ is the quark field. The indices 
take the following ranges $1$~$\leq$~$a$~$\leq$~$N_A$,
$1$~$\leq$~$I$~$\leq$~$N_F$ and $1$~$\leq$~$i$~$\leq$~$\Nf$ where $N_F$ and 
$N_A$ are the dimensions of the fundamental and adjoint representations 
respectively, $\Nf$ is the number of quark flavours and $f^{abc}$ are the 
colour group structure constants. The covariant derivatives which determine
$G^a_{\mu\nu}$ are 
\begin{equation} 
D_\mu c^a ~=~ \partial_\mu c^a ~-~ g f^{abc} A^b_\mu c^c ~~,~~ 
D_\mu \psi^{iI} ~=~ \partial_\mu \psi^{iI} ~+~ i g T^a A^a_\mu \psi^{iI} ~.  
\end{equation} 
In \cite{12}, the LCO formalism was derived which involves an additional scalar
field $\sigma$ which is related to the dimension two composite operator
$\half A_\mu^2$. The relevant Lagrangian is 
\begin{equation}
L^{\mbox{\footnotesize{LCO}}} ~=~ L^{\mbox{\footnotesize{QCD}}} ~-~ 
\frac{\sigma^2}{2g^2 \zeta(g)} ~+~ \frac{1}{2 g \zeta(g)} \sigma A^a_\mu 
A^{a \, \mu} ~-~ \frac{1}{8\zeta(g)} \left( A^a_\mu A^{a \, \mu} \right)^2 
\end{equation}
where there is an extra contribution to the quartic gluon interaction and
$L^{\mbox{\footnotesize{LCO}}}$ contains the usual covariant gauge fixing terms
though we will only consider the Landau gauge case, $\alpha$~$=$~$0$. The 
quantity $\zeta(g)$ is a function of the coupling constant which has been 
computed to $O(g^2)$ in the Landau gauge in \cite{12,13,14} and is such that it
ensures the generating functional underlying the formalism satisfies a 
homogeneous renormalization group equation, \cite{12}. For this article we note
that the relevant terms are 
\begin{eqnarray}
\frac{1}{g^2\zeta(g)} &=& \left[ \frac{( 13 C_A - 8 T_F \Nf )}{9N_A} 
\right. \nonumber \\
&& \left. +~ \left( 2685464 C_A^3 T_F \Nf - 1391845 C_A^4 
- 213408 C_A^2 C_F T_F \Nf - 1901760 C_A^2 T_F^2 \Nf^2 
\right. \right. \nonumber \\
&& \left. \left. ~~~~~ 
+~ 221184 C_A C_F T_F^2 \Nf^2 + 584192 C_A \Nf^3 T_F^3 
- 55296 C_F T_F^3 \Nf^3 
\right. \right. \nonumber \\
&& \left. \left. ~~~~~ 
-~ 65536 T_F^4 \Nf^4 \right) \frac{g^2}{5184 \pi^2 N_A (35 C_A-16 T_F \Nf) 
(19 C_A-8 T_F \Nf)} \right] ~.  
\end{eqnarray}
In \cite{12,13,14} the $\sigma$ field develops a non-zero vacuum expectation 
value when one computes the one loop effective potential of $\sigma$ which is
\begin{eqnarray}
V(\sigma) &=& 
\frac{9N_A}{2} \lambda_1 \sigma^{\prime \, 2} \nonumber \\
&& +~ \left[ \frac{3}{64} \ln \left( \frac{g \sigma^\prime}{\mu^2} \right)
+ C_A \left(
-~ \frac{351}{8} C_F \lambda_1 \lambda_2
+ \frac{351}{16} C_F \lambda_1 \lambda_3
- \frac{249}{128} \lambda_2
+ \frac{27}{64} \lambda_3
\right)
\right. \nonumber \\
&& \left. ~~~~~ 
+~ C_A^2 \left(
-~ \frac{81}{16} \lambda_1 \lambda_2
+ \frac{81}{32} \lambda_1 \lambda_3
\right)
\right. \nonumber \\
&& \left. ~~~~~ 
+ \left(
-~ \frac{13}{128}
- \frac{207}{32} C_F \lambda_2
+ \frac{117}{32} C_F \lambda_3
\right)
\right] \frac{g^2 N_A \sigma^{\prime \, 2}}{\pi^2} ~+~ O(g^4)  
\label{effpot} 
\end{eqnarray} 
where we have set 
\begin{equation}
\lambda_1 ~=~ [13 C_A-8 T_F \Nf]^{-1} ~~,~~
\lambda_2 ~=~ [35 C_A-16 T_F \Nf]^{-1} ~~,~~
\lambda_3 ~=~ [19 C_A-8 T_F \Nf]^{-1} ~~,~~ 
\end{equation}
\begin{equation}
\sigma ~=~ \frac{9 N_A}{(13 C_A - 8 T_F \Nf)} \sigma^\prime 
\end{equation} 
and $\mu$ is the usual $\MSbar$ renormalization scale, which is introduced to 
retain a dimensionless coupling constant in dimensional regularization. 

Now we consider the relation between the pole mass and the running gluon mass
in the Curci-Ferrari model, \cite{25}, which includes the BRST invariant mass 
operator 
\begin{equation}
L^{\mbox{\footnotesize{mass}}} ~=~ \frac{1}{2} m^2 A^a_\mu A^{a \, \mu} ~-~
\alpha m^2 \bar{c}^a c^a 
\end{equation}
where $m$ is the bare mass. With this term the gluon and ghost propagators in 
the Landau gauge are
\begin{equation}
-~ \frac{\delta^{ab}}{(k^2-m^2)} \left[ \eta_{\mu\nu} ~-~ 
\frac{k_\mu k_\nu}{k^2} \right] ~~~,~~~ \frac{\delta^{ab}}{k^2}
\end{equation}
respectively. With these it is a straightforward exercise to compute the one
loop correction to the gluon two-point function. In this respect the one loop  
snail diagram derived from the quartic gluon interaction cannot be neglected
in the massive case. The result of our computation for the pole mass in the
Curci-Ferrari model is
\begin{eqnarray}
m^2_{\mbox{\footnotesize{CF}}} &=& \left[ 1 ~+~ \left( \left( 
\frac{313}{576} ~-~ \frac{35}{192} \ln \left( \frac{m^2(\mu)}{\mu^2} 
\right) ~-~ \frac{11\pi\sqrt{3}}{128} \right) C_A \right. \right. \nonumber \\
&& \left. \left. ~~~~~~~~~~~~+~ \left( \frac{1}{12} \ln \left( 
\frac{m^2(\mu)}{\mu^2} \right) - \frac{5}{36} \right) T_F \Nf \right) 
\frac{g^2}{\pi^2} ~+~ O(g^4) \right] m^2(\mu)  
\label{mpole}
\end{eqnarray}
where $m_{\mbox{\footnotesize{o}}}$ $=$ $m(\mu) Z_m$ is the bare mass, $m(\mu)$
is the running mass and $\mu$ is the renormalization mass scale. We have
renormalized with the usual one loop $\MSbar$ renormalization constants. As a
check on the expression, we note that it reduces to the same relation given
in \cite{29} when $\Nf$ $=$ $0$. Moreover, we have verified the expression 
of \cite{29} for arbitrary $\alpha$ prior to specifying the Landau gauge which
provided a non-trivial check on the symbolic manipulation programmes we used
for this article.  

We have repeated the above computation for the LCO Lagrangian where the bare
mass is now defined to be $\sigma/[g\zeta(g)]$, \cite{12,13,14}, which at 
leading order is  
\begin{equation}
m^2_{\mbox{\footnotesize{o}}} ~=~ \frac{(13C_A - 8T_F\Nf)}{9 N_A} g \sigma ~. 
\end{equation}
With the additional interactions the expression for the LCO pole mass is of a
similar form 
\begin{eqnarray}
m^2_{\mbox{\footnotesize{LCO}}} &=& \left[ 1 ~+~ \left( \left( 
\frac{287}{576} ~-~ \frac{3}{64} \ln \left( \frac{m^2(\mu)}{\mu^2} 
\right) ~-~ \frac{11\pi\sqrt{3}}{128} \right) C_A \right. \right. \nonumber \\
&& \left. \left. ~~~~~~~~~~~~-~ \frac{1}{9} T_F \Nf \right) 
\frac{g^2}{\pi^2} ~+~ O(g^4) \right] m^2(\mu)  
\label{mpolelco} 
\end{eqnarray}
for massless quarks in the Landau gauge. Equipped with this result we can now
extend the method of \cite{12,13,14} for estimating a gluon mass. In 
\cite{12,14} the minimum of the effective potential (\ref{effpot}) was 
determined by solving $\frac{d V(\sigma)}{d \sigma}$ $=$ $0$. Since the factors
multiplying the classical effective mass are coupling constant dependent, this 
is equivalent to extremizing 
$V^{\mbox{\footnotesize{eff}}}(m^2_{\mbox{\footnotesize{o}}})$. 
However, it seems to us that an alternative approach is to compute instead the 
extremum of $V^{\mbox{\footnotesize{eff}}}(m^2_{\mbox{\footnotesize{LCO}}})$ 
where one inverts (\ref{mpolelco}) to obtain $m(\mu)$ as a function of 
$m^2_{\mbox{\footnotesize{LCO}}}$ and then substitutes this into 
(\ref{effpot}). Thus we find
\begin{eqnarray}
V^{\mbox{\footnotesize{eff}}} \left(m^2_{\mbox{\footnotesize{LCO}}}\right) &=& 
\left[ \frac{9}{2} \lambda_1 + \left( - \frac{29}{128} - \frac{207}{32}C_F 
\lambda_2 + \frac{117}{32}C_F \lambda_3 \right. \right. \nonumber \\
&& \left. \left. ~+~ C_A \left( - \frac{351}{8}C_F\lambda_1 \lambda_2
+\frac{351}{16}C_F \lambda_1 \lambda_3 - \frac{183}{64} \lambda_1
\right. \right. \right. \nonumber \\
&& \left. \left. \left. ~~~~~~~~~~~~-~ \frac{249}{128} \lambda_2 
+ \frac{27}{64} \lambda_3 + \frac{99}{128}\pi \sqrt{3} \lambda_1 \right) 
\right. \right. \nonumber \\ 
&& \left. \left. ~+~ C_A^2 \left( - \frac{81}{16} \lambda_1 \lambda_2
+ \frac{81}{32} \lambda_1 \lambda_3 \right) + \frac{3}{64} \ln \left( 
\frac{m^2_{\mbox{\footnotesize{LCO}}}}{\mu^2} \right) \right. \right. 
\nonumber \\
&& \left. \left. ~+~ \frac{27}{64}C_A \lambda_1 \ln \left( 
\frac{m^2_{\mbox{\footnotesize{LCO}}}}{\mu^2} \right) 
\right) \frac{g^2}{\pi^2} \right] 
\frac{\left( 13 C_A - 8 T_F \Nf \right)^2}{81 N_{\! A}} g^2 \zeta^2(g) 
m^4_{\mbox{\footnotesize{LCO}}} 
\end{eqnarray}
Repeating the process to find a minimum necessitates solving  
\begin{eqnarray}
0 &=& \left[ \frac{9}{2} \lambda_1 + \left( - \frac{13}{64} - \frac{207}{32}C_F 
\lambda_2 + \frac{117}{32}C_F \lambda_3 \right. \right. \nonumber \\
&& \left. \left. ~+~ C_A \left( - \frac{351}{8}C_F\lambda_1 \lambda_2
+\frac{351}{16}C_F \lambda_1 \lambda_3 - \frac{339}{128} \lambda_1
\right. \right. \right. \nonumber \\
&& \left. \left. \left. ~~~~~~~~~~~~-~ \frac{249}{128} \lambda_2 
+ \frac{27}{64} \lambda_3 + \frac{99}{128}\pi \sqrt{3} \lambda_1 \right) 
\right. \right. \nonumber \\ 
&& \left. \left. ~+~ C_A^2 \left( - \frac{81}{16} \lambda_1 \lambda_2
+ \frac{81}{32} \lambda_1 \lambda_3 \right) + \frac{3}{64} \ln \left( 
\frac{m^2_{\mbox{\footnotesize{LCO}}}}{\mu^2} \right) \right. \right. 
\nonumber \\
&& \left. \left. ~+~ \frac{27}{64}C_A \lambda_1 \ln \left( 
\frac{m^2_{\mbox{\footnotesize{LCO}}}}{\mu^2} \right) 
\right) \frac{g^2}{\pi^2} \right] 
\frac{\left( 13 C_A - 8 T_F \Nf \right)^2}{81 N_{\! A}} g^2 \zeta^2(g) 
m^4_{\mbox{\footnotesize{LCO}}} 
\label{mincond}
\end{eqnarray}
which corresponds to the condition
\begin{equation}
\frac{d V\left(m^2_{\mbox{\footnotesize{LCO}}}\right)} 
{d m^2_{\mbox{\footnotesize{LCO}}}} ~=~ 0 ~.  
\end{equation} 
We have not in fact substituted for the explicit expression for $\zeta(g)$ 
since this function factorizes off the expression for the location of the 
minimum. If we were to include that part of the series which was already known 
it would introduce an unnecessary truncation error into our final estimates for
the pole mass. At this point to solve for the mass a scale needs to be chosen 
for $\mu$. In \cite{12,13,14}, the choice of scale was such that it removed the
logarithm terms. For this potential we will take a more general approach and 
instead set $m^2_{\mbox{\footnotesize{LCO}}}$~$=$~$s \mu^2$ where $s$ is an 
arbitrary parameter. This means we have determined an equation for the value of
the coupling constant as a function of $s$. In other words
\begin{eqnarray} 
y &=& 36 C_A \left( 16 T_F \Nf - 35 C_A \right) \left[ \left( 3465 \pi \sqrt{3}
+ 4620 \ln(s) - 25690 \right) C_A^2 - 864 C_F T_F \Nf \right. \nonumber \\
&& \left. ~~~~~~~~~~~~~~~~~~~~~~~~~~~~~~~~+~ \left( 19240 - 1584 \pi \sqrt{3} 
- 3792 \ln(s) \right) C_A T_F \Nf \right. \nonumber \\
&& \left. ~~~~~~~~~~~~~~~~~~~~~~~~~~~~~~~~+~ \left( 768 \ln(s) - 3328 \right) 
T_F^2 \Nf^2 \right]^{-1}
\end{eqnarray}  
where $y$ $=$ $C_A g^2/(16\pi^2)$. Through the definition of the running 
coupling constant we have the one loop relation 
\begin{equation} 
\frac{g^2(\mu)}{16\pi^2} ~=~ \left[ \beta_0 \ln \left[ \frac{\mu^2}
{\Lambda^2_{\mbox{\footnotesize{$\MSbar$}}}}\right] \right]^{-1}  
\end{equation} 
where 
\begin{equation} 
\beta_0 ~=~ \frac{11}{3} C_A ~-~ \frac{4}{3} T_F \Nf ~. 
\end{equation} 
Hence, we can relate the coupling constant to the scale $\mu$ and 
$\Lambda_{\mbox{\footnotesize{$\MSbar$}}}$ and deduce a value for 
$m^2_{\mbox{\footnotesize{LCO}}}$. We find
\begin{eqnarray}
m_{\mbox{\footnotesize{LCO}}} &=& 
\Lambda^{(\Nf)}_{\mbox{\footnotesize{$\MSbar$}}} \exp \left[ \,-\, \left( 
\left( 3465 \pi \sqrt{3} - 25690 \right) C_A^2 - 864 C_F T_F \Nf \right. 
\right. \nonumber \\
&& \left. \left. ~~~~~~~~~~~~~~~~~~~+~ \left( 19240 - 1584 \pi \sqrt{3} \right) 
C_A T_F \Nf ~-~ 3328 T_F^2 \Nf^2 \right) \right. \nonumber \\
&& \left. ~~~~~~~~~~~~~~ \left( \frac{}{} \! 24 \left( 11 C_A - 4 T_F \Nf 
\right) \left( 35 C_A - 16 T_F \Nf \right) \right)^{-1} \right] 
\label{massform}
\end{eqnarray}
which is the main result of this article. It turns out that this is 
{\em independent} of the intermediate parameter $s$. In other words, no matter
what scale $\mu$ one chooses, one will always obtain the same value for the
solution to (\ref{mincond}) at one loop.
{\begin{table}[ht] 
\begin{center} 
\begin{tabular}{|c||c|c|} 
\hline
$N_{\!f}$ & $m_{\mbox{\footnotesize{$SU(2)$}}}/\Lambda^{(\Nf)}_{\mbox{\footnotesize{$\MSbar$}}}$ & 
$m_{\mbox{\footnotesize{$SU(3)$}}}/\Lambda^{(\Nf)}_{\mbox{\footnotesize{$\MSbar$}}}$ \\ 
\hline
 0 & 2.10 & 2.10 \\ 
 2 & 1.54 & 1.74 \\ 
 3 & 1.24 & 1.55 \\ 
\hline
\end{tabular} 
\end{center} 
\begin{center} 
{Table 1. One loop estimates of the gluon effective mass for $SU(2)$ and
$SU(3)$.}
\end{center} 
\end{table}}  

We have given the explicit values of the pole mass estimates from 
(\ref{massform}) for $SU(2)$ and $SU(3)$ in Table~1. Compared with the 
classical effective gluon mass estimates of \cite{12,13,14} the Yang-Mills 
estimates have increased by about $5\%$ for $SU(3)$. However, 
for $\Nf$~$\neq$~$0$ there is a significant decrease. Although this is 
disappointing it is important to recognise that since they have been derived in
a scale independent and therefore renormalization group invariant manner, they 
may be closer to the true result, though the inclusion of quark mass may alter 
these estimates. 

We conclude with several remarks. First, we have constructed a one loop
renormalization group invariant pole mass for the gluon using the LCO effective
potential of \cite{12,13,14}. However, it would be interesting to see whether
this feature persists at the next order. This only requires an extension of the
present one loop result since the two loop LCO effective potential is 
available. Although we have ignored quark mass effects it seems that if one
could include quark condensates in the LCO formalism in addition to that for
$\half A^{a\,\mu} A^a_\mu$ then it might be possible to ascertain the extent
to which condensates could be responsible for the quark and gluon masses. If
the renormalization scale invariance persists even at one loop for this 
scenario then one would not have to worry about solving a multi-scale type
renormalization group equation. Our final comment concerns the situation where
a gluon mass is dynamically generated through, say, the LCO formalism. If this 
is the case then one would have to include additional contributions due to a 
gluon mass to the existing quark pole mass multi-loop estimates.  

\vspace{1cm} 
\noindent
{\bf Acknowledgement.} This work was supported in part by {\sc PPARC} through a
research studentship, (REB). We also thank Prof D.R.T. Jones and Dr C. McNeile 
for discussions.

\end{document}